\documentclass[lettersize,journal]{IEEEtran}

\usepackage{cite}
\usepackage{amsmath,amssymb,amsfonts}
\usepackage{algorithm}    % For algorithm environment
\usepackage{algpseudocode}
\usepackage{graphicx}     % Only one copy needed
\usepackage{textcomp}
\usepackage{xcolor}
\usepackage{url}
\usepackage{times}
\usepackage{booktabs}
\usepackage{float}

\def\BibTeX{{\rm B\kern-.05em{\sc i\kern-.025em b}\kern-.08em
    T\kern-.1667em\lower.7ex\hbox{E}\kern-.125emX}}

\begin{document}

\title{RoGBot: Relationship-Oblivious Graph-based
Neural Network with Contextual Knowledge for Bot Detection}

\author{
    \IEEEauthorblockN{Ashutosh Anshul\IEEEauthorrefmark{1}, Mohammad Zia Ur Rehman\IEEEauthorrefmark{1},  Sri Akash Kadali\IEEEauthorrefmark{2}, Nagendra Kumar\IEEEauthorrefmark{1}\IEEEauthorrefmark{3} \thanks{\IEEEauthorrefmark{3}Corresponding Author \\ All the authors have equal contribution}}
    \\
    \IEEEauthorblockA{\IEEEauthorrefmark{1}Indian Institute of Technology Indore \IEEEauthorrefmark{2}University of Maryland, College Park, USA
    \\ashutoshanshul01@gmail.com, phd2101201005@iiti.ac.in, kadali18@umd.edu, nagendra@iiti.ac.in}
}

\maketitle

\begin{abstract}
Detecting automated accounts (bots) among genuine users on platforms like Twitter remains a challenging task due to the evolving behaviors and adaptive strategies of such accounts. While recent methods have achieved strong detection performance by combining text, metadata, and user relationship information within graph-based frameworks, many of these models heavily depend on explicit user–user relationship data. This reliance limits their applicability in scenarios where such information is unavailable. To address this limitation, we propose a novel multimodal framework that integrates detailed textual features with enriched user metadata while employing graph-based reasoning without requiring follower–following data. Our method uses transformer-based models (e.g., BERT) to extract deep semantic embeddings from tweets, which are aggregated using max pooling to form comprehensive user-level representations. These are further combined with auxiliary behavioral features and passed through a GraphSAGE model to capture both local and global patterns in user behavior. Experimental results on the Cresci-15, Cresci-17, and PAN 2019 datasets demonstrate the robustness of our approach, achieving accuracies of 99.8\%, 99.1\%, and 96.8\%, respectively, and highlighting its effectiveness against increasingly sophisticated bot strategies.

\end{abstract}

\begin{IEEEkeywords} 
GraphSAGE, Semantic Embeddings, Bot Detection, Representation Learning, Social Networks
\end{IEEEkeywords}

\section{Introduction}

Social media platforms have become essential tools for communication, sharing news, and fostering public engagement. However, the increasing presence of automated accounts, commonly known as bots, has raised serious concerns. These bots are often designed to mimic human behavior, allowing them to spread misinformation, artificially boost engagement, and distort public discussions. This can have serious consequences, especially during sensitive events, such as political elections or health emergencies, where the integrity of information is critical.

Earlier approaches to bot detection focused mainly on building classical machine learning methods by leveraging user profile features~\cite{yang2022botometer, yang2020scalable}, and text behavioral patterns~\cite{cresci2016dna, lee2013warningbird}. While earlier methods were effective in identifying simple bots, they struggled with more advanced bots that carefully mimicked human users. To address these limitations, researchers began exploring deep learning techniques, including long short-term memory (LSTM)~\cite{kudugunta2018deep} networks, bi-directional LSTM~\cite{wei2019twitter} networks, and transformers~\cite{dukic2020you}, to model tweet content. More recently, Graph Neural Networks (GNNs)~\cite{ali2019detect, feng2021botrgcn, feng2022heterogeneity} have gained attention for their ability to incorporate user interactions and relationships, improving detection performance.

\begin{figure}[t]
\centering
% \captionsetup{justification=centering}
\includegraphics[width=0.5\textwidth]{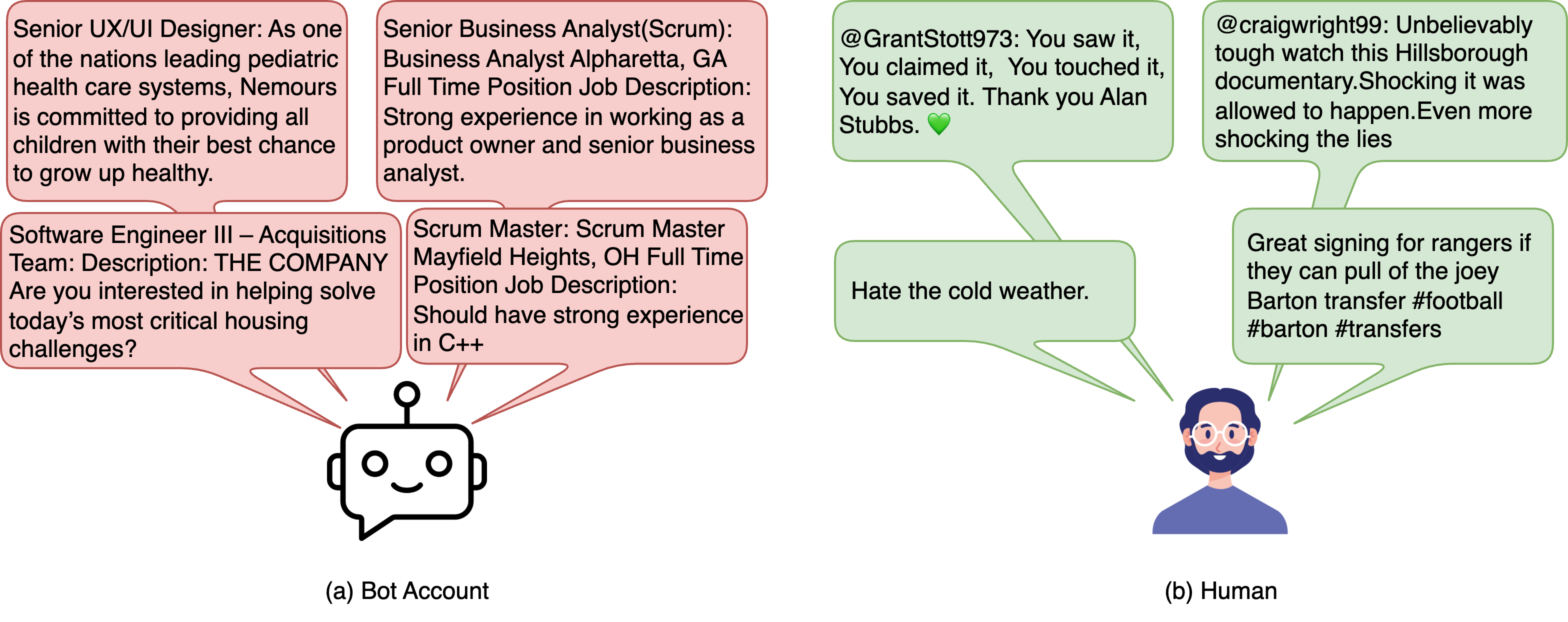} % Replace with your actual image file name
\caption{Difference between tweets by (a) a bot (red) and (b) a human (green). Notice how the content of bot tweets is similar in content, length, and use of entities such as emojis and hashtags. On the contrary, human tweets are not very consistent in terms of content similarity, length, and the use of emojis and hashtags.}
\label{fig:example1}
\end{figure}

The release of datasets such as Cresci-15~\cite{cresci2015fame} and TwiBot-20~\cite{feng2021twibot} has enabled a new direction in bot detection research. These datasets include not only user tweets and metadata but also detailed follower–following relationships. This has allowed the development of graph-based models like DMIBot and HHG-Bot, which use the social graph to model higher-order interactions and improve classification accuracy. However, two major challenges remain

\begin{itemize}
    \item  Modern bots have evolved to behave more like real users. They generate human-like language, mimic natural interaction patterns, and often maintain realistic-looking profiles. This makes them harder to distinguish from genuine accounts using static features alone. As a result, detection models need to go beyond simple behavior-based or text-based cues and incorporate more adaptive and context-aware reasoning.
    \item Due to recent restrictions on Twitter’s API access and the high cost of manual labeling, many available datasets like Cresci-17~\cite{cresci2017paradigm} and PAN 2019~\cite{daelemans2019overview} lack follower–following information entirely. For a dataset to have this follower–following information, it may be difficult to reach an optimal balance where the dataset is large enough to be representative yet densely connected enough to support meaningful graph-based reasoning. While datasets like Cresci-15~\cite{cresci2015fame} and TwiBot-20~\cite{feng2021twibot} provide relatively rich relationship structures, such datasets remain limited and are not easily replicable. Hence, in datasets like Cresci-17~\cite{cresci2017paradigm} and PAN 2019~\cite{daelemans2019overview}, where user relationships are missing, graph-based methods that rely on social connections become less effective.
\end{itemize}

To address these issues, we propose a method that leverages user metadata and tweet-level features to construct an initial user–user interaction graph. A graph neural network is then applied to update user-level representations for the final task of bot detection. Unlike traditional machine learning methods, our approach uses deep learning to extract tweet features, which are subsequently processed through a graph neural network for classification. This graph-based architecture enables the model to capture complex relationships among users, enhancing its reasoning ability and achieving state-of-the-art performance on the Cresci-17~\cite{cresci2017paradigm} and PAN 2019~\cite{daelemans2019overview} datasets. Furthermore, unlike recent graph-based approaches such as DMIBot~\cite{lin2025dmibot} and HHG-Bot~\cite{wang2025hhg}, our method constructs the user–user interaction graph without relying on follower–following data, making it applicable to datasets where such information is unavailable. A very high-level overview of the pipeline is illustrated in Figure~\ref{fig:example}. In summary:
\begin{enumerate}
    \item We propose an end-to-end bot detection framework that uses a graph-based approach to model user-to-user relationships by combining features from user metadata and tweet content. This multimodal integration enables effective cross-user feature interaction, where both behavioral patterns and semantic information from tweets contribute to the creation of node features and edges, resulting in a more informative and expressive graph structure.
    \item In contrast to relying on follower–following relationships to build user interaction graphs, we propose to construct the user–user graph based solely on behavioral and metadata similarities between users. This design choice allows the model to be applied to datasets such as Cresci-17~\cite{cresci2017paradigm} and PAN 2019~\cite{daelemans2019overview}, where explicit social network information is unavailable, broadening its applicability to a wider range of real-world scenarios.
    \item To enable graph-based reasoning without reliance on social links, we leverage GraphSAGE~\cite{hamilton2017inductive} to aggregate information across behaviorally similar users. This allows the model to capture both local and higher-order interactions, improving its ability to generalize across diverse bot behavior patterns.
    \item We evaluate our method using widely accepted bot detection benchmarks and compare it with state-of-the-art models. Results from both performance and ablation studies show that our approach outperforms baseline models. Notably, our model reaches comparable performance to state-of-the-art approaches even without using the follower–following relationships information.
\end{enumerate}

\section{Related Works}
The detection of social bots, automated accounts masquerading as real users, has evolved significantly due to their increasing sophistication and impact on online discourse. Early approaches predominantly relied on shallow machine learning models with handcrafted features such as posting frequency, account age, or friend–follower ratios. However, as bots have grown more human-like and capable of mimicking nuanced behavior, traditional detection techniques have struggled to remain effective. Consequently, recent research has shifted toward more robust paradigms that incorporate contextual embeddings, graph structures, and self-supervised learning frameworks. This section surveys foundational and recent advancements in social bot detection, with a focus on graph-based, transformer-based, and interpretable unsupervised methods.

Cresci et al. \cite{cresci2017paradigm} were among the first to demonstrate the limitations of conventional detection systems in the face of evolving bot behavior. Their work introduced the concept of social spambots, which convincingly emulate human activity through coordinated group behavior, realistic tweeting patterns, and polished user profiles. By benchmarking a variety of detection strategies—including Twitter’s internal tools and human annotations—they revealed a significant performance gap in identifying this new bot archetype. Importantly, their study advocated for group-level behavioral analysis, laying the groundwork for bot detection approaches that leverage collective behavior rather than isolated user signals.

Building on the need for better feature representation, Lopez-Joya et al. \cite{lopez2024exploring} conducted a large-scale empirical study comparing account-based, content-based, and hybrid feature sets for bot detection. They found that traditional machine learning models, such as Random Forests, can outperform deep learning baselines when provided with well-engineered features. Their hybrid approach, which included metadata (e.g., follower count, profile completeness), stylistic content patterns, and novel features such as profile color binning and content compression ratios, showed robust generalization across datasets. Their work reaffirms the value of interpretable and domain-specific feature engineering in bot detection.

Recent advances have also leveraged transformer-based models to capture nuanced language patterns and social signals~\cite{rehman2025implihatevid,rehman2025multimodal,bansal2025retrieval}. Sallah et al. \cite{sallah2024fine} developed a framework that fine-tunes models like BERT, RoBERTa, and GPT-3 to detect bots at the tweet level. Their approach significantly outperformed traditional word embedding baselines, incorporating SHAP values and LLM-based explanations to enhance model transparency. This work demonstrated that, when adapted for social media data, transformer architectures can excel at detecting bots while maintaining interpretability.

To address the limitations of supervised approaches that require labeled data, Wang et al. \cite{wang2024unsupervised} proposed BotDCGC, an unsupervised, graph-based method that combines graph autoencoders, deep clustering, and contrastive learning. Their framework constructs user graphs using profile and activity features, then applies a graph attention encoder to learn discriminative embeddings. BotDCGC’s ability to adapt to unseen bot behaviors and imbalanced datasets highlights the growing interest in label-free, structure-aware detection techniques.

Further advancing this trend, Yang et al. \cite{yang2024sebot} introduced SeBot, a contrastive learning framework that incorporates structural entropy minimization and heterophily-aware aggregation to detect bots in a self-supervised manner. By constructing community-aware encoding trees and customizing message-passing beyond homophily assumptions, SeBot excels at identifying bots that intentionally connect with humans to camouflage their identity. This approach demonstrated state-of-the-art results on large-scale datasets like TwiBot-20 and MGTAB, emphasizing the potential of multi-view graph modeling for resilient detection.

Complementing these efforts, Peng et al. \cite{peng2024unsupervised} proposed UnDBot, an unsupervised and interpretable framework rooted in structural information theory. Instead of relying on interaction graphs or opaque neural networks, UnDBot models user similarity through a multi-relational graph that captures posting behavior, influence, and follow-to-follower ratios. It then minimizes heterogeneous structural entropy to derive a hierarchical encoding tree for user clustering and employs stationary distribution and entropy-based scoring for community labeling. Experiments across multiple benchmark datasets show that UnDBot not only achieves competitive accuracy but also offers strong interpretability, making it suitable for real-world deployment.

Some existing studies detect bots by focusing on one of the following information: user profile metadata, tweet text, or follower–following relationships. For example, SGBot~\cite{yang2020scalable} uses readily available metadata such as statuses\_count, followers\_count, and friends\_count, along with derived features like tweet frequency, followers growth rate, and friends growth rate. It also shows that selecting a carefully chosen subset of training data leads to better generalization than simply merging all data. Wei et al.~\cite{wei2019twitter} apply a Bidirectional LSTM on word embeddings generated using GloVe~\cite{pennington2014glove}. HOFA~\cite{ye2023hofa} leverages graph-based user–user relationships and introduces a homophily-oriented augmentation module together with a frequency-adaptive attention module to handle heterophilous disguised bots, which many graph-based models miss. While effective, these approaches rely on a single modality of information, despite other valuable sources, such as user profile data and tweet text, being readily available.

To overcome this limitation, other studies combine two or more types of features. BotRGCN~\cite{feng2021botrgcn}, for example, integrates both user profile data and user–user relationships in a graph-based framework. Similarly, BotBuster~\cite{ng2023botbuster} and Kudugunta et al.~\cite{kudugunta2018deep} combine tweet text with user metadata. More recently, DMIBot~\cite{lin2025dmibot} incorporates all three sources of information independently and then fuses them through a Synergistic Interaction Module to make the final prediction. Although these multimodal, graph-based approaches achieve strong performance, most of them construct graphs using user–user relationship information, which is not always available in real-world datasets. Our model addresses this issue by building graphs based on similarity in textual and metadata features instead. This makes it more adaptable to cases where follower–following information is missing, while still benefiting from the reasoning power of graph-based methods.

\begin{figure*}[h]
\centering
\includegraphics[width=1\textwidth]{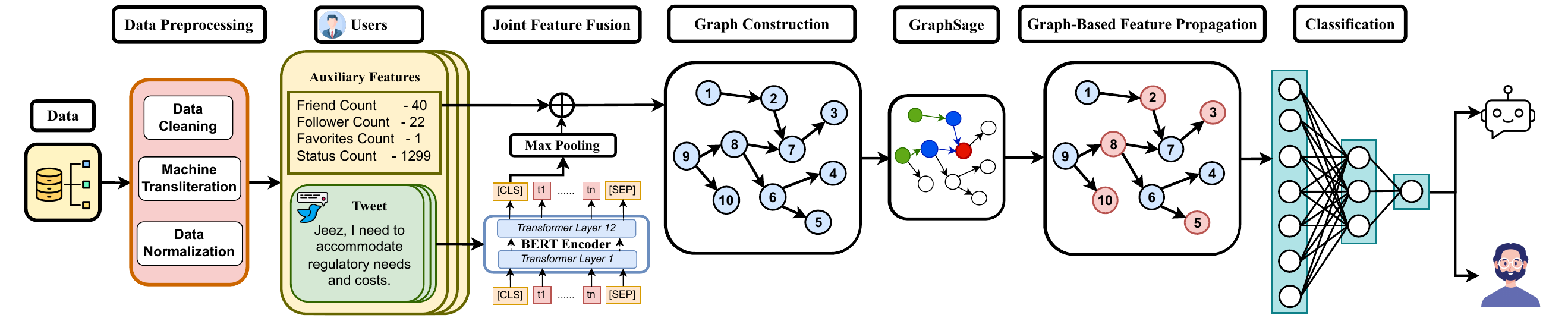} % Replace with your actual image file name
\caption{\textbf{System Architecture:} We integrate tweet-level semantic embeddings (via BERT) with user metadata to form a unified representation, which we then use to construct a user–user graph based on feature similarity. We apply GraphSAGE to refine node representations through neighborhood aggregation, and finally, we use a classification layer to predict whether a user is a bot.}
\label{fig:architecture}
\end{figure*}

\section{Methodology}
 
To address the limitations of existing bot detection models that rely heavily on social network structures, we propose a graph-based framework that utilizes only tweet text and user-level (auxiliary) features. This design avoids dependence on follower-following information, making the method more adaptable and applicable in settings where such data is unavailable or incomplete.

The proposed model, as illustrated in Figure~\ref{fig:architecture}, begins with data preprocessing, where raw tweet texts are cleaned to improve the quality of downstream feature extraction. Next, auxiliary feature extraction collects user-level metadata that captures behavioral patterns relevant to bot identification. In parallel, BERT~\cite{devlin2019bert}-based feature extraction is used to obtain contextual embeddings for each tweet, which are then aggregated to form a single semantic representation per user.

These two components are then combined in the feature fusion step, where the aggregated tweet embeddings and auxiliary features are concatenated into a unified vector. Based on similarity between these fused features, the graph construction phase generates a user-user adjacency matrix, enabling the discovery of implicit relationships without relying on explicit network links.

Finally, deep learning classification is performed using a GraphSage~\cite{hamilton2017inductive} Network, which refines user representations by incorporating information from neighboring nodes in the graph. The GraphSage output is passed through fully connected layers to produce the final predictions.

Graph neural networks are particularly suited for this task due to their ability to model complex dependencies and perform relational reasoning. Thus, by leveraging only user-level and tweet-based features, our approach enables effective bot detection within a graph framework, without the need for explicit social connections.

In the following subsections, we provide a detailed description of each stage.

%%%%%%%%%%%%%%%%%%%%%%%%%%%%%%%%%%%%%%%%%%%%%%%%%%%%%%%%%%%%%%%%%%%%%%%%%%%%%%%
\subsection{Data Preprocessing}

Preprocessing is a critical step in any text-based machine learning pipeline, particularly in social media analysis, where the data is inherently noisy and inconsistent. Tweets often contain irregular spelling, excessive punctuation, emojis, non-standard scripts, and informal language. If left unprocessed, these inconsistencies can degrade the quality of learned representations and negatively impact model performance. Therefore, effective preprocessing is necessary to standardize the input, reduce noise, and enhance the robustness of feature extraction methods, especially when using pretrained language models like BERT~\cite{devlin2019bert}.

Let the collection of raw tweets from all users be denoted as  
\begin{equation}
    T = \{ t_{k,i} \mid k = 1,\ldots,N; \; i = 1,\ldots,n_k \}
\end{equation}
where \( t_{k,i} \) represents the \( i \)th tweet of user \( u_k \), and \( n_k \) is the number of tweets posted by user \( u_k \). Each tweet \( t_{k,i} \) undergoes a sequence of preprocessing operations, defined as follows:

\begin{itemize}
    \item \textbf{Cleaning:} The first step involves removing extra white spaces, punctuation, and any erroneous or non-standard characters from the raw tweet text. This is formally represented as:  
    \begin{equation}
        \tilde{t}_{k,i} = \text{Clean}(t_{k,i}), \quad \forall \, t_{k,i} \in T
    \end{equation}
    \item \textbf{Transliteration:} Tweets written in non-Roman scripts (e.g., Devanagari, Arabic, etc.) are converted to Roman script to ensure compatibility with language models trained predominantly on Romanized text. A transliteration function \( T(\cdot) \) is applied: 
    \begin{equation}
        t^{*}_{k,i} = T(\tilde{t}_{k,i}), \quad \forall \, \tilde{t}_{k,i} \in T
    \end{equation}
    \item \textbf{Normalization:} This step includes multiple transformations: converting emojis to their textual descriptions, converting all characters to lowercase, and applying other text normalization techniques to maintain consistency across the dataset. The transformation function is denoted by:
    \begin{equation}
        \hat{t}_{k,i} = \text{Transform}(t^{*}_{k,i}), \quad \forall \, t^{*}_{k,i} \in T
    \end{equation}
\end{itemize}

The final preprocessed set of tweets \( \{ \hat{t}_{k,i} \} \) serves as input for subsequent feature extraction, ensuring that the language model processes clean, standardized, and semantically meaningful text.

%%%%%%%%%%%%%%%%%%%%%%%%%%%%%%%%%%%%%%%%%%%%%%%%%%%%%%%%%%%%%%%%%%%%%%%%%%%%%%%
\subsection{Auxiliary Feature Extraction}

In addition to textual content, user-level metadata provides valuable behavioral cues that can significantly aid in distinguishing between genuine users and automated accounts. Unlike tweet text, which captures content semantics, auxiliary features reflect a user's overall activity patterns, interaction tendencies, and account-level behavior. Prior work in social bot detection has shown that such features often exhibit distinct statistical patterns for bot accounts compared to human users—for instance, bots may follow large numbers of accounts with minimal reciprocal engagement, or post an unusually high volume of content within short time spans.

To capture such behavioral signals, we extract a set of auxiliary features for each user \( u_k \), denoted as  
\begin{equation}
    \mathbf{a}_k = \Bigl[ f^{(1)}_k, \, f^{(2)}_k, \, f^{(3)}_k, \, f^{(4)}_k \Bigr] \in \mathbb{R}^4
\end{equation}
where each component represents a specific user profile attribute:
\begin{align*}
f^{(1)}_k &= \text{Followers Count},\\[1mm]
f^{(2)}_k &= \text{Friends Count},\\[1mm]
f^{(3)}_k &= \text{Statuses Count},\\[1mm]
f^{(4)}_k &= \text{Favorites Count}.
\end{align*}

These features are chosen due to their general availability and their potential to reflect discrepancies in social behavior. For example, a high follower-to-following ratio may indicate influence or bot-like popularity inflation, while unusually high tweet or favorite counts may signal automation. Importantly, these features are lightweight to extract and do not rely on full access to the user’s network graph, making them well-suited for scalable and privacy-conscious detection frameworks.

%%%%%%%%%%%%%%%%%%%%%%%%%%%%%%%%%%%%%%%%%%%%%%%%%%%%%%%%%%%%%%%%%%%%%%%%%%%%%%%
\subsection{BERT Feature Extraction and Aggregation}

The textual content of user posts is one of the most crucial sources of information for identifying bots, as it directly reflects patterns in language use, topical focus, and potential automation cues. Bots often exhibit repetitive phrasing, unnatural text generation, or highly uniform topics—all of which can be effectively captured through modern language models. To exploit these nuances, we employ a pre-trained BERT~\cite{devlin2019bert} model for tweet-level feature extraction.

BERT (Bidirectional Encoder Representations from Transformers) is a deep contextual language model that captures a sentence's left and right context through transformer-based self-attention mechanisms~\cite{kasu2025dhumordarkhumorunderstanding}. For each preprocessed tweet \( \hat{t}_{k,i} \) from user \( u_k \), the BERT model generates a dense embedding via its pooler output:
\begin{equation}
    \mathbf{v}_{k,i} = f_{\text{BERT}}(\hat{t}_{k,i}) \in \mathbb{R}^d
\end{equation}

Since users can post varying numbers of tweets, we aggregate these tweet-level embeddings to derive a single representation \( \mathbf{v}_k \) for each user. Two common strategies are considered:
\begin{itemize}
    \item \textbf{Average Pooling:} This computes the mean representation across all tweets:
    \begin{equation}
        \mathbf{v}_k = \frac{1}{n_k} \sum_{i=1}^{n_k} \mathbf{v}_{k,i}
    \end{equation}
    \item \textbf{Max Pooling:} This selects the maximum value across all tweet embeddings along each dimension:
    \begin{equation}
        \mathbf{v}_k = \max_{1 \leq i \leq n_k} \mathbf{v}_{k,i}
    \end{equation}
\end{itemize}

These aggregation strategies enable the model to generate consistent user-level representations regardless of tweet count, while preserving the rich semantic information learned from individual tweets. The resulting embeddings are then used in subsequent stages for feature fusion and graph construction.

%%%%%%%%%%%%%%%%%%%%%%%%%%%%%%%%%%%%%%%%%%%%%%%%%%%%%%%%%%%%%%%%%%%%%%%%%%%%%%%
\subsection{Joint Feature Fusion}

To create a unified representation that leverages both semantic and behavioral information, we fuse the aggregated tweet embeddings with the auxiliary user-level features. While the BERT-based embeddings \( \mathbf{v}_k \) capture linguistic and contextual patterns from the user's tweets, the auxiliary features \( \mathbf{a}_k \) encode behavioral characteristics from profile metadata. Combining these complementary views helps the model form a more holistic understanding of each user.

The fusion is performed via vector concatenation, resulting in a joint feature vector for each user \( u_k \):
\begin{equation}
    \mathbf{f}_k = \Bigl[ \mathbf{v}_k;\ \mathbf{a}_k \Bigr] \in \mathbb{R}^{d+4}
\end{equation}
Thus, for a total of \( N \) users, we obtain the set of fused feature vectors:
\begin{equation}
    F = \{ \mathbf{f}_1, \mathbf{f}_2, \ldots, \mathbf{f}_N \}
\end{equation}

These joint features serve as the input node representations in the subsequent graph-based learning phase, enabling the model to reason over content and profile-level cues.

%%%%%%%%%%%%%%%%%%%%%%%%%%%%%%%%%%%%%%%%%%%%%%%%%%%%%%%%%%%%%%%%%%%%%%%%%%%%%%%

\begin{figure*}[h]
\centering
\includegraphics[width=0.65\textwidth]{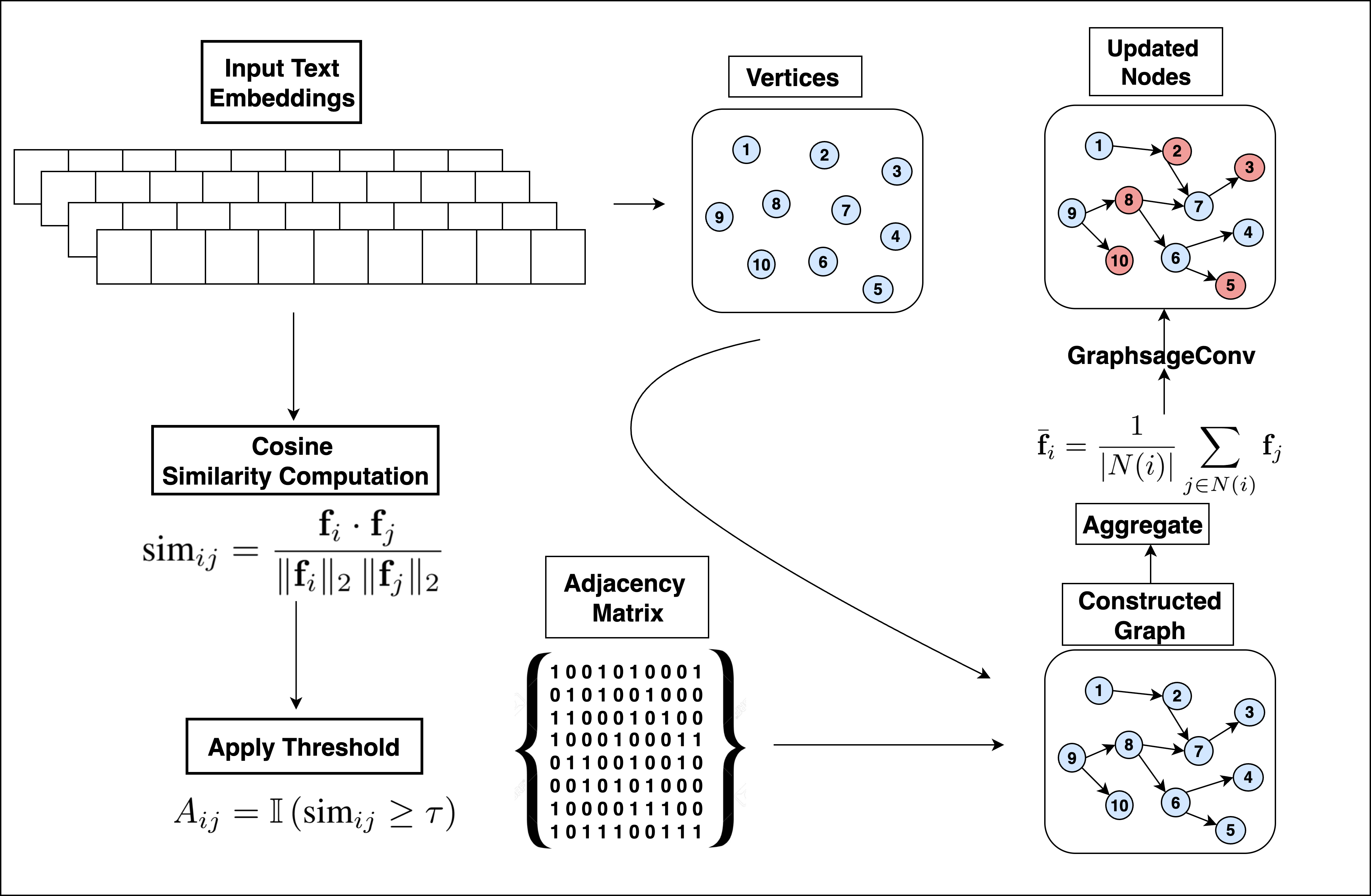} % Replace with your actual image file name
\caption{\textbf{Graph Construction:} We construct a user graph by representing each user as a node with fused textual and metadata features. We compute pairwise cosine similarity and add edges only when the similarity exceeds a threshold, ensuring balanced connectivity and sparsity. This lets us capture meaningful semantic relations without relying on follower-following links.}
\label{fig:graphcons}
\end{figure*}
\subsection{Graph Construction}

Graph-based models~\cite{feng2021botrgcn, ye2023hofa, feng2021satar} have demonstrated strong reasoning capabilities across various domains, including social bot detection, due to their ability to capture complex interactions among entities. To leverage this strength, we formulate the problem within a graph-based learning framework. However, unlike prior approaches that rely heavily on follower-following relationships to establish user connections, our method constructs the graph using only tweet-level semantics and user-level metadata. This choice addresses two key challenges: (i) follower-following data may not always be accessible or complete, and (ii) such dependency may result in either overly sparse or overly dense subgraphs, limiting scalability and generalization.

In our approach, each user \( u_k \) is represented as a node in the graph, initialized with the joint feature vector \( \mathbf{f}_k \) that fuses textual and auxiliary features. To define edges between users, we compute pairwise cosine similarity between these feature vectors. Intuitively, users (whether bots or humans) with similar behavioral or linguistic patterns are likely to be semantically connected, and cosine similarity provides an effective and scalable metric to capture this relation.

Formally, we define the undirected graph \( G = (V, E) \), where each node \( v_k \in V \) corresponds to a user. The cosine similarity between two users \( u_i \) and \( u_j \) is given by:
\begin{equation}
    \text{sim}_{ij} = \frac{\mathbf{f}_i \cdot \mathbf{f}_j}{\|\mathbf{f}_i\|_2 \, \|\mathbf{f}_j\|_2}
\end{equation}
An edge is established between two users if their similarity exceeds a predefined threshold \( \tau \):
\begin{equation}
    A_{ij} = \mathbb{I} \left( \text{sim}_{ij} \geq \tau \right)
\end{equation}
Here, \( A \in \mathbb{R}^{N \times N} \) is the adjacency matrix, and \( \mathbb{I}(\cdot) \) is the indicator function. The threshold \( \tau \) is selected empirically to balance graph connectivity and sparsity. For clarity, the graph construction algorithm is summarized in Algorithm \ref{alg:graph_construction}

Through this formulation, we construct a graph purely from accessible features, enabling scalable and generalizable learning without reliance on explicit social links. This graph, along with the node features, serves as input to the graph neural network for downstream classification.

\begin{algorithm}[ht]
\caption{Graph Construction Using Cosine Similarity}
\label{alg:graph_construction}
\begin{algorithmic}[1]
\State \textbf{Input:} Joint feature matrix \( F \in \mathbb{R}^{N \times (d+4)} \), threshold \( \tau \)
\State \textbf{Output:} Adjacency matrix \( A \in \{0,1\}^{N \times N} \)
\For{$i = 1$ to $N$}
    \For{$j = i+1$ to $N$}
        \State Compute $ \text{sim}_{ij} = \frac{\mathbf{f}_i \cdot \mathbf{f}_j}{\|\mathbf{f}_i\|_2 \, \|\mathbf{f}_j\|_2} $
        \If{$ \text{sim}_{ij} \geq \tau $}
            \State Set $ A_{ij} = A_{ji} = 1 $
        \Else
            \State Set $ A_{ij} = A_{ji} = 0 $
        \EndIf
    \EndFor
\EndFor
\State \Return $ A $
\end{algorithmic}
\end{algorithm}

%%%%%%%%%%%%%%%%%%%%%%%%%%%%%%%%%%%%%%%%%%%%%%%%%%%%%%%%%%%%%%%%%%%%%%%%%%%%%%%

\subsection{Graph-Based Feature Propagation (GraphSAGE)}  

To enable the learning of user representations that incorporate both individual and neighborhood-level information, we employ the GraphSAGE~\cite{hamilton2017inductive} architecture for feature propagation. Unlike classical transductive models such as Graph Convolutional Networks (GCNs)~\cite{kipf2016semi}, GraphSAGE~\cite{hamilton2017inductive} is designed for inductive learning, allowing generalization to unseen nodes or users, which is crucial in dynamic environments like social media where new users continuously emerge.

For a given user node \( u_i \) with initial feature vector \( \mathbf{f}_i \), GraphSAGE~\cite{hamilton2017inductive} generates an updated representation by aggregating information from its local neighborhood \( N(i) \), as defined by the previously constructed adjacency matrix. The neighborhood features are first averaged to obtain:
\begin{equation}
    \bar{\mathbf{f}}_i = \frac{1}{|N(i)|} \sum_{j \in N(i)} \mathbf{f}_j
\end{equation}
This aggregated representation is then concatenated with the node’s own features and passed through a non-linear transformation:
\begin{equation}
    \mathbf{h}_i^{(1)} = \sigma\Bigl( W^{(1)} \cdot \text{concat}\bigl(\mathbf{f}_i, \bar{\mathbf{f}}_i\bigr) + b^{(1)} \Bigr)
\end{equation}
where \( W^{(1)} \) and \( b^{(1)} \) are learnable parameters, and \( \sigma(\cdot) \) is a non-linear activation function such as ReLU.

This mechanism allows each node to update its representation by incorporating signals from semantically or behaviorally similar users—those with whom it shares high cosine similarity in the joint feature space. By doing so, the model captures higher-order correlations that may be indicative of bot-like patterns, such as coordinated posting behavior or shared metadata signatures.

The decision to use cosine similarity for graph construction plays a key role here, as it ensures that feature propagation occurs between users exhibiting similar behavior or linguistic patterns. This encourages the GraphSAGE model to refine user embeddings in a way that amplifies discriminative traits relevant to bot detection.

The final output of the GraphSAGE~\cite{hamilton2017inductive} layers is a user embedding \( \mathbf{h}_i^{(L)} \) that encodes both the user’s own characteristics and contextual cues from its surrounding neighborhood in the graph, serving as a robust input for downstream classification.

%%%%%%%%%%%%%%%%%%%%%%%%%%%%%%%%%%%%%%%%%%%%%%%%%%%%%%%%%%%%%%%%%%%%%%%%%%%%%%%%

\subsection{Deep Learning Classification}

Once the feature propagation step using GraphSAGE~\cite{hamilton2017inductive} is complete, each user node \( u_i \) now has an updated representation \( \mathbf{h}_i^{(1)} \), which integrates both its own features and contextual information from its neighbors. These enriched embeddings are passed through a series of fully connected layers to perform the final classification.

\paragraph{Fully Connected Layers}  
To capture higher-order interactions and non-linear patterns, the representation \( \mathbf{h}_i^{(1)} \) is passed through multiple fully connected (FC) layers. For each layer \( l \in \{2, \ldots, L\} \), the transformation is defined as:
\begin{equation}
    \mathbf{h}_i^{(l)} = \sigma\Bigl( W^{(l)} \mathbf{h}_i^{(l-1)} + b^{(l)} \Bigr)
\end{equation}
where \( W^{(l)} \) and \( b^{(l)} \) are learnable parameters, and \( \sigma(\cdot) \) is a non-linear activation function such as ReLU.

To improve generalization and stabilize training, batch normalization is applied after each FC layer:
\begin{equation}
    \hat{\mathbf{h}}_i^{(l)} = \text{BN} \bigl( \mathbf{h}_i^{(l)} \bigr)
\end{equation}
followed by dropout to prevent overfitting:
\begin{equation}
    \tilde{\mathbf{h}}_i^{(l)} = \text{Dropout}\bigl( \hat{\mathbf{h}}_i^{(l)},\; p \bigr)
\end{equation}
where \( p \) is the dropout rate.

\paragraph{Output Layer}  
The output from the final FC layer \( \tilde{\mathbf{h}}_i^{(L)} \) is passed through a linear transformation followed by a softmax function to obtain the predicted class probabilities:
\begin{equation}
    \mathbf{z}_i = W_{\text{out}} \tilde{\mathbf{h}}_i^{(L)} + b_{\text{out}}
\end{equation}
\begin{equation}
    \hat{\mathbf{y}}_i = \text{softmax}(\mathbf{z}_i)
\end{equation}
Here, \( \hat{\mathbf{y}}_i \in \mathbb{R}^C \) represents the predicted probability distribution over the \( C \) classes for user \( u_i \).

\paragraph{Loss Function}  
To train the model, we minimize the categorical cross-entropy loss over all labeled users:
\begin{equation}
    \mathcal{L} = - \frac{1}{N} \sum_{i=1}^{N} \sum_{c=1}^{C} y_{i,c} \log \hat{y}_{i,c}
\end{equation}
where \( y_{i,c} \) is the one-hot encoded ground truth label for user \( u_i \), and \( \hat{y}_{i,c} \) is the predicted probability for class \( c \).

%%%%%%%%%%%%%%%%%%%%%%%%%%%%%%%%%%%%%%%%%%%%%%%%%%%%%%%%%%%%%%%%%%%%%%%%%%%%%%%
In summary, our method combines tweet content and user-level metadata to construct informative user representations using a pre-trained BERT model and feature concatenation. These representations are used to build a graph based on cosine similarity, capturing relational patterns without relying on follower-following information. We then apply a GraphSAGE~\cite{hamilton2017inductive} model to propagate features across this graph, followed by fully connected layers to perform the final classification of users as bots or humans.

\begin{table*}[htbp]
\centering
\caption{\textbf{Comparison Results on Cresci-15:} We report Accuracy and F1-Score on the Cresci-15 dataset. \textbf{Relationship} represents that the model uses the neighborhood or follower-following information. Best result under each metric is highlighted in \textbf{bold}. }
\label{table:cresci15_comparison}
\begin{tabular}{|l|c|c|c|c|c|c|c|}
\hline
\textbf{Method} & \textbf{Text} & \textbf{Metadata} & \textbf{Relationship} & \textbf{Graph-Based} & \textbf{Accuracy} & \textbf{F1-Score} \\ \hline
Botometer~\cite{yang2022botometer} & & \checkmark & & & 57.90 & 66.90 \\ \hline
SGBot~\cite{yang2020scalable} & & \checkmark & & & 77.10 & 77.91 \\ \hline
Wei et al.~\cite{wei2019twitter} & \checkmark & & & & 96.10 & 82.65 \\ \hline
HOFA~\cite{ye2023hofa} & & & \checkmark & \checkmark & 97.52 & 98.06 \\ \hline
Graph-Hist~\cite{magelinski2020graph} & & & \checkmark & \checkmark & 77.38 & 84.47 \\ \hline
BotRGCN~\cite{feng2021botrgcn} & & \checkmark & \checkmark & \checkmark & 96.50 & 97.30 \\ \hline
BotBuster~\cite{ng2023botbuster} & \checkmark & \checkmark & & & 96.90 & 97.53 \\ \hline
Kudugunta et al.~\cite{kudugunta2018deep} & \checkmark & \checkmark & & & 96.11 & 96.88 \\ \hline
RGT~\cite{feng2022heterogeneity} & \checkmark & & \checkmark & \checkmark & 97.15 & 97.78 \\ \hline
SATAR~\cite{feng2021satar} & \checkmark & \checkmark & \checkmark & \checkmark & 93.40 & 95.05 \\ \hline
DMIBot~\cite{lin2025dmibot} & \checkmark & \checkmark & \checkmark & \checkmark & 97.84 & 98.12 \\ \hline
\textbf{RoGBot} & \checkmark & \checkmark & & \checkmark & \textbf{99.79} & \textbf{98.51} \\ \hline
\end{tabular}
\end{table*}

\section{Experiments and Results}

\subsection{Datasets Used}
This project utilizes the following datasets:
\begin{itemize}
    \item \textbf{Cresci-15}~\cite{cresci2015fame}: We begin our evaluation using the Cresci-15 dataset, a widely adopted benchmark for bot detection. This dataset includes 3,900 Twitter accounts, evenly split between 1,950 genuine human users and 1,950 fake follower accounts. Cresci-15 offers rich multimodal information, comprising user profile attributes, tweet text, and follower-following relationships, which supports the development and comparison of models that integrate metadata, language features, and graph-based reasoning. This allows for a fair comparison with recent graph-based approaches that rely on neighborhood information to model user behavior.

    \item \textbf{Cresci-17}~\cite{cresci2017paradigm}: We use the Cresci-2017 dataset to evaluate our approach to bot detection on Twitter. This dataset includes 11,017 Twitter accounts, comprising both genuine human users and a diverse set of spambots. The bot accounts are categorized into two types: social spambots and traditional spambots. Among these, social spambots are particularly challenging to detect due to their sophisticated behavior and human-like profiles, including detailed bios, profile pictures, and realistic activity patterns. To focus on this more difficult and realistic detection scenario, we use only genuine user accounts and social spambots for training and evaluation, comprising 8386 accounts.
    
    \item \textbf{PAN 2019}~\cite{daelemans2019overview}: We also evaluate our approach using the PAN 2019 Author Profiling dataset, which was released as part of the PAN shared task on bots and gender profiling. The dataset consists of Twitter user accounts, each represented by a collection of tweets authored by that user. Each account is labeled either as a bot or a human, enabling binary classification. The dataset covers multiple languages, but in our study, we focus on the English subset for consistency. Compared to earlier datasets, PAN 2019 includes more recent accounts and is designed to reflect real-world bot detection challenges, with bots exhibiting both automated and semi-automated behavior. The dataset provides tweet collections for every user without requiring follower-following information, making it well-suited for our graph construction approach based solely on textual and metadata features. Although it lacks detailed user-level features, it offers valuable insights into the textual characteristics of user-generated content, making it particularly useful for analyzing linguistic patterns associated with bots.
    
\end{itemize}

The combination of these datasets provides a diverse set of features and benchmarks, ensuring a robust evaluation of the proposed framework.

\subsection{Comparison Results on Cresci-15}

To assess the overall robustness of our model, particularly its similarity-based graph construction, we evaluate it on the Cresci-15~\cite{cresci2015fame} dataset, which includes user metadata, tweet texts, and follower–following information for each user. We report the model’s performance in terms of accuracy and F1-score, and compare it with recent works in Table~\ref{table:cresci15_comparison}. The columns ``Text," ``Metadata," and ``Relationship" indicate the modalities used by each model, corresponding to tweet text, user metadata, and follower–following information, respectively. The ``Graph-Based" column indicates whether the model employs a graph-based reasoning architecture. Generally, graph-based models incorporate relationship information to build the graph. Our approach is distinct in that it uses a graph-based method without relying on follower–following data. The stronger performance of multimodal models like BotRGCN~\cite{feng2021botrgcn} and BotBuster~\cite{ng2023botbuster}, compared to single-modality models such as Botometer~\cite{yang2022botometer} and SGBot~\cite{yang2020scalable}, highlights the advantage of integrating multiple sources of information. HOFA~\cite{ye2023hofa}, which performs well using only relationship data, further demonstrates the power of graph-based reasoning. However, the superior performance of DMIBot~\cite{lin2025dmibot} and our model over HOFA~\cite{ye2023hofa} can be attributed to their multimodal design. Notably, our model outperforms DMIBot~\cite{lin2025dmibot} by 1.95\% in accuracy and 0.39\% in F1-score, despite not using relationship data, underscoring the robustness of our graph construction based on feature cosine similarity. This also enhances the model's applicability to scenarios where relationship information is unavailable.

%%%%%%%%%%%%%%%%%%%%%%%%%%%%%%%%%%%%%%%%%%%%%%%%%%%%%%%%%%%%%%%%%%%%%%%%%%%%%%%
\subsection{Comparison Results on Cresci-17}

\begin{table*}[htbp]
\centering
\caption{\textbf{Comparison Results on Cresci-17}: We report Accuracy, Precision, Recall and F1-Score achived my models on the Cresci-17 dataset. Best results are highlighted in \textbf{bold}. For the comparison methods, we report the average score obtained across the Mixed1 and Mixed2 sets.}
\label{table:mib_comparison}
\begin{tabular}{|l|c|c|c|c|c|c|c|}
\hline
\textbf{Method}           & \textbf{Accuracy} & \textbf{Precision} & \textbf{Recall} & \textbf{F1-Score} \\ \hline
Digital DNA Compression~\cite{pasricha2019detecting} & 98.1 & \textbf{99.2} & 96.8 & \textbf{98.0} \\ \hline
K-common Substring - Supervised~\cite{cresci2017social} & 97.4 & 98.0 & \textbf{96.9} & 97.4 \\ \hline
Yang et al.~\cite{yang2013empirical} & 56.8 & 64.5 & 29.0 & 39.3 \\ \hline
Miller et al.~\cite{miller2014twitter} & 50.4 & 51.1 & 33.2 & 40.3 \\ \hline
Ahmed et al.~\cite{ahmed2013generic} & 93.3 & 92.9 & 94.0 & 93.4 \\ \hline
\textbf{RoGBot}          & \textbf{99.1} & 98.9 & 95.5 & 97.2 \\ \hline
\end{tabular}
\end{table*}

\begin{table*}[htbp]
\centering
\caption{\textbf{Comparison Results on PAN 2019 Dataset:} Accuracy, Precision, Recall and F1-Score achieved on PAN 2019 dataset. Best result under each metric is highlighted in \textbf{bold}.}
\label{table:pan_comparison}
\begin{tabular}{|l|c|c|c|c|c|c|c|}
\hline
\textbf{Method} & \textbf{Accuracy} & \textbf{Precision} & \textbf{Recall} & \textbf{F1-Score} \\ \hline
PAN 2019 Overview               & 96.0 & 95.0 & 95.5 & 95.3 \\ \hline
% Semantic and Syntactic Features & 92.0 & 91.5 & 91.8 & 91.7 \\ \hline
Random Forest Classifier        & 84.0 & 83.5 & 83.0 & 83.3 \\ \hline
MLPs for Author Profiling       & 90.5 & 90.5 & 90.0 & 90.3 \\ \hline
BiLSTM for Bots Detection       & 93.5 & 93.2 & 93.0 & 93.1 \\ \hline
Hybrid SVM and LSTM             & 94.5 & 94.3 & 94.0 & 94.2 \\ \hline
\textbf{RoGBot}                 & \textbf{96.8} & \textbf{96.5} & \textbf{96.9} & \textbf{96.8} \\ \hline
\end{tabular}
\end{table*}
\subsection{Comparison Results on PAN Dataset}

To demonstrate the applicability of our model in scenarios where relationship information is unavailable, we evaluate it on the Cresci-17~\cite{cresci2017paradigm} dataset, which contains only user metadata and tweet text. We compare our results with existing approaches, as shown in Table~\ref{table:mib_comparison}. Our model achieves a test accuracy of 99.1\%, indicating strong overall performance on this dataset. While the Digital DNA Compression method shows slightly better results in some other metrics, our model remains comparable and demonstrates robust detection capabilities.

%%%%%%%%%%%%%%%%%%%%%%%%%%%%%%%%%%%%%%%%%%%%%%%%%%%%%%%%%%%%%%%%%%%%%%%%%%%%%%%

We further evaluate our model in a setting where neither relationship information nor user metadata is available. To simulate this, we test the model on the PAN 2019~\cite{daelemans2019overview} dataset, which contains 100 tweets per user but lacks any user metadata. In this setup, we omit metadata features when defining user representations and constructing the initial graph. Our model achieves a test accuracy of 96.8\% as reported in Table~\ref{table:pan_comparison}, outperforming existing methods such as BiLSTM and the Hybrid SVM-LSTM. The balance between precision and recall across both classes indicates that our method effectively reduces both false positives and false negatives. Compared to models that rely solely on textual features, our use of graph-based learning and auxiliary signals offers a clear performance advantage.

%%%%%%%%%%%%%%%%%%%%%%%%%%%%%%%%%%%%%%%%%%%%%%%%%%%%%%%%%%%%%%%%%%%%%%%%%%%%%%%
\subsection{Performance Gain Analysis}

\begin{table*}[htbp]
\centering
\caption{\textbf{Ablation Study Results on Cresci-17:} We ablate on the contribution of individual user metadata features and the GraphSAGE on the performance of the model.}
\label{tab:ablation_results}
\begin{tabular}{|l|c|c|c|c|c|c|c|}
\hline
\textbf{Method} & \textbf{Accuracy} & \textbf{Precision} & \textbf{Recall} & \textbf{F1-Score} \\ \hline
Without Status Count       & 95.0 & 93.5 & 91.0 & 92.2 \\ \hline
Without Followers Count    & 97.0 & 94.8 & 94.0 & 94.4 \\ \hline
Without Friends Count      & 96.0 & 94.0 & 92.0 & 93.0 \\ \hline
Without Favorites Count    & 93.5 & 91.1 & 90.1 & 90.5 \\ \hline
Without GraphSAGE          & 94.5 & 92.5 & 90.5 & 91.4 \\ \hline
\textbf{RoGBot}            & \textbf{99.1} & \textbf{98.9} & \textbf{95.5} & \textbf{97.2} \\ \hline
\end{tabular}
\end{table*}

We aim to evaluate the impact of auxiliary features and the GraphSAGE component on the overall performance of our model. To do this, we adopt a leave-one-out strategy, where each component is removed individually to assess its contribution. The model is then evaluated using the remaining components, and the results are reported in Table~\ref{tab:ablation_results}. The ablation study shows that removing the GraphSAGE~\cite{hamilton2017inductive} component reduces accuracy from 99.1\% to 94.5\%, highlighting its importance in capturing relational patterns among users. Similarly, excluding auxiliary features such as Status Count and Followers Count results in significant drops in both precision and recall, especially for bot detection. These results confirm that the integration of semantic, auxiliary, and graph-based features is critical for achieving robust detection performance.

%%%%%%%%%%%%%%%%%%%%%%%%%%%%%%%%%%%%%%%%%%%%%%%%%%%%%%%%%%%%%%%%%%%%%%%%%%%%%%%

\subsection{t-SNE Visualization of Test Embeddings}
\begin{figure}[h!]
    \centering
    \includegraphics[width=0.4\textwidth]{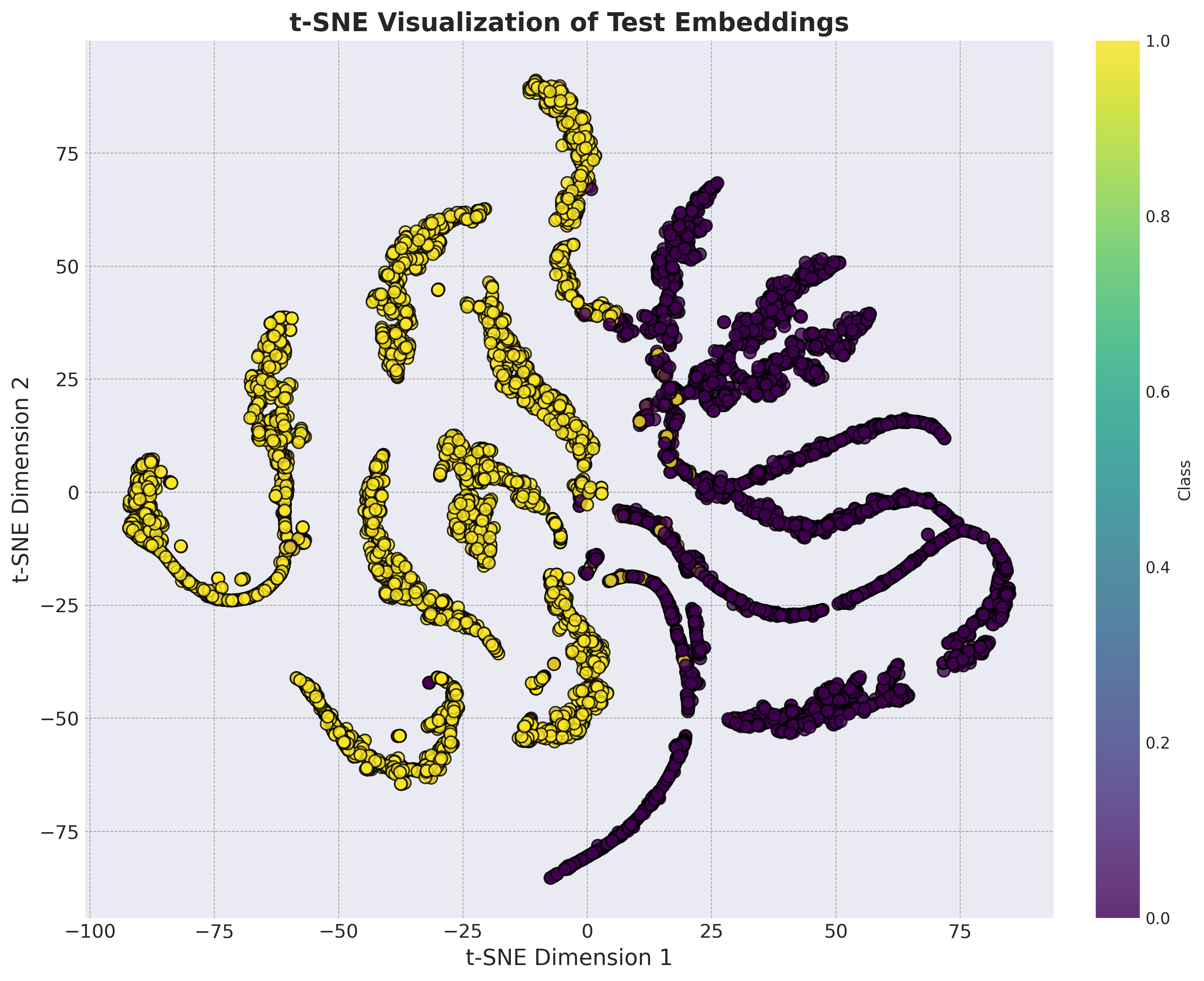}
    \caption{t-SNE visualization of learned embeddings for bot and human users}
    \label{fig:tsne}
\end{figure}

The t-SNE plot illustrates how effectively the model distinguishes between human and bot users in the embedding space. Red clusters represent bot users, while blue clusters denote human users. The clear separation between the two indicates that the model successfully captures behavioral patterns that differentiate the classes. Minor overlaps may reflect users with borderline behavior or ambiguous posting activity.

\subsection{Impact of Similarity Threshold on Accuracy}
\begin{figure}[h!]
    \centering
    \includegraphics[width=0.4\textwidth]{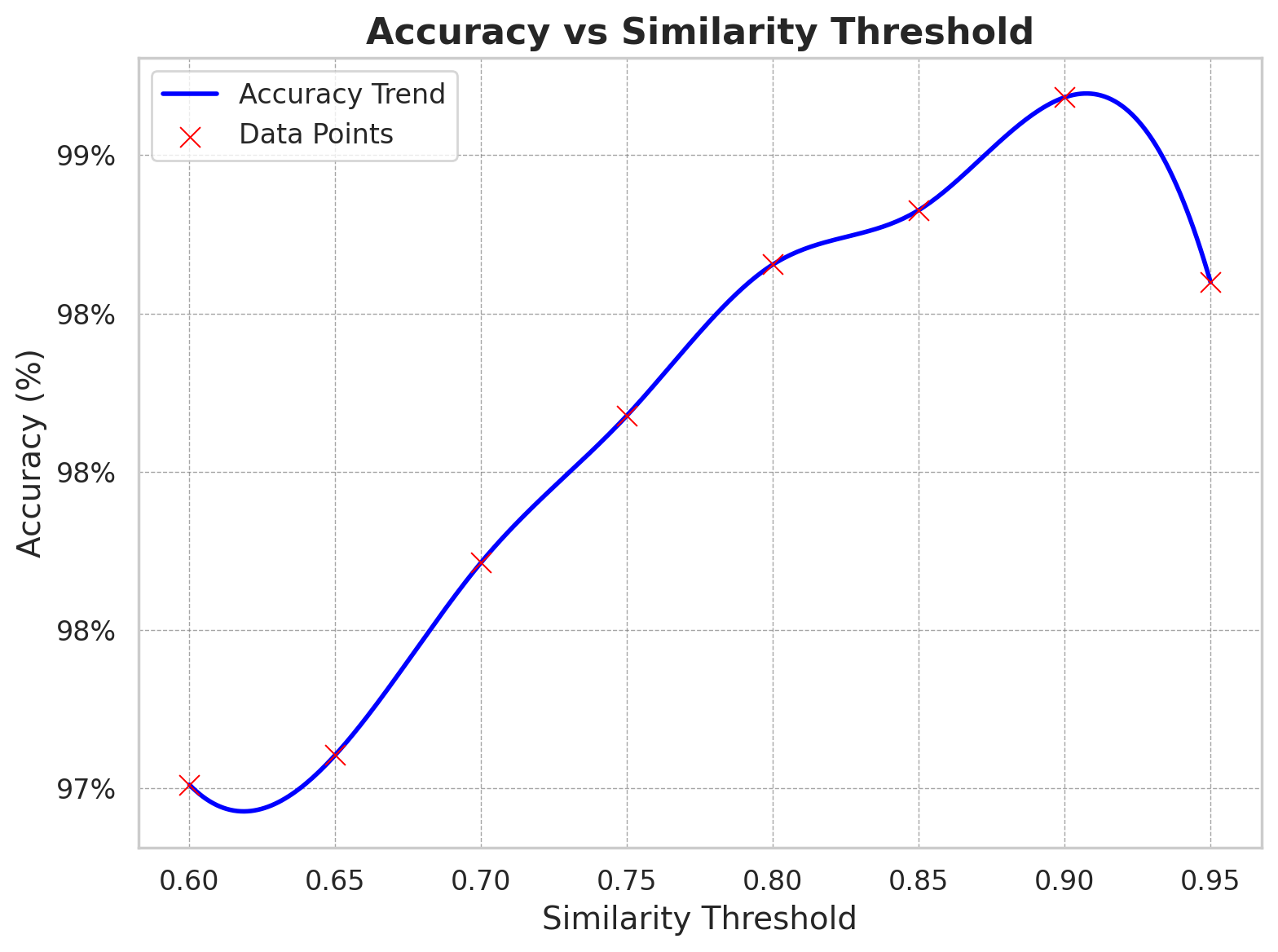}
    \caption{Accuracy trend as a function of similarity threshold}
    \label{fig:threshold}
\end{figure}

The similarity threshold plays a critical role in defining graph-based relationships among users. As shown in Fig.~\ref{fig:threshold}, accuracy improves with increasing threshold values, reaching above 99\% when the threshold is set to 0.90. Beyond this point, accuracy slightly declines, suggesting that overly strict similarity criteria may remove meaningful connections, weakening the graph’s ability to capture relevant user interactions. This emphasizes the importance of fine-tuning the threshold to balance graph sparsity and connectivity. An optimal threshold ensures that the graph retains both strong intra-class cohesion and sufficient inter-class separation, which is essential for accurate classification.

\subsection{Precision-Recall Curve}
\begin{figure}[h!]
    \centering
    \includegraphics[width=0.5\textwidth]{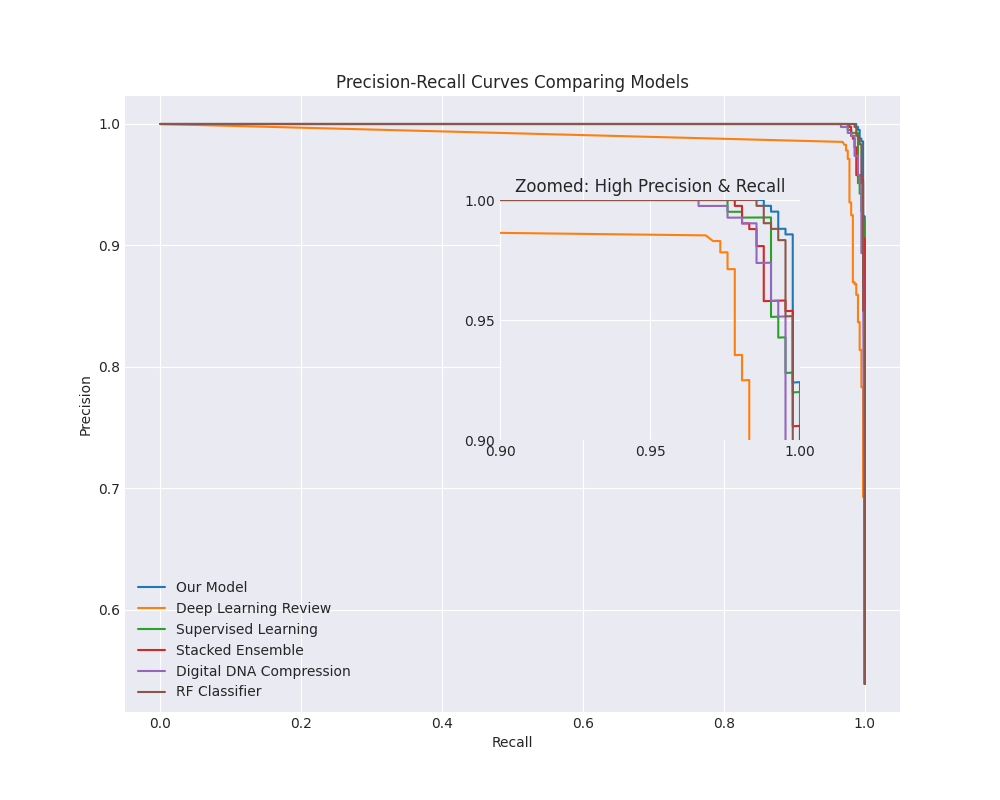}
    \caption{Precision-Recall curves comparing our model with existing methods}
    \label{fig:pr_curve}
\end{figure}

The Precision-Recall (PR) curve offers valuable insights into the model’s ability to balance precision and recall. The x-axis denotes recall (true positive rate), while the y-axis represents precision. In Fig.~\ref{fig:pr_curve}, our model (blue curve) consistently maintains higher precision across a wide range of recall values, effectively minimizing false positives while preserving high true positive rates. The steep and elevated shape of the PR curve reflects the model’s strong ability to distinguish between bots and human users. This indicates a reliable detection system that can perform well even under class imbalance, which is often common in real-world bot detection scenarios.

\subsection{ROC Curve (Receiver Operating Characteristic)}
\begin{figure}[h!]
    \centering
    \includegraphics[width=0.5\textwidth]{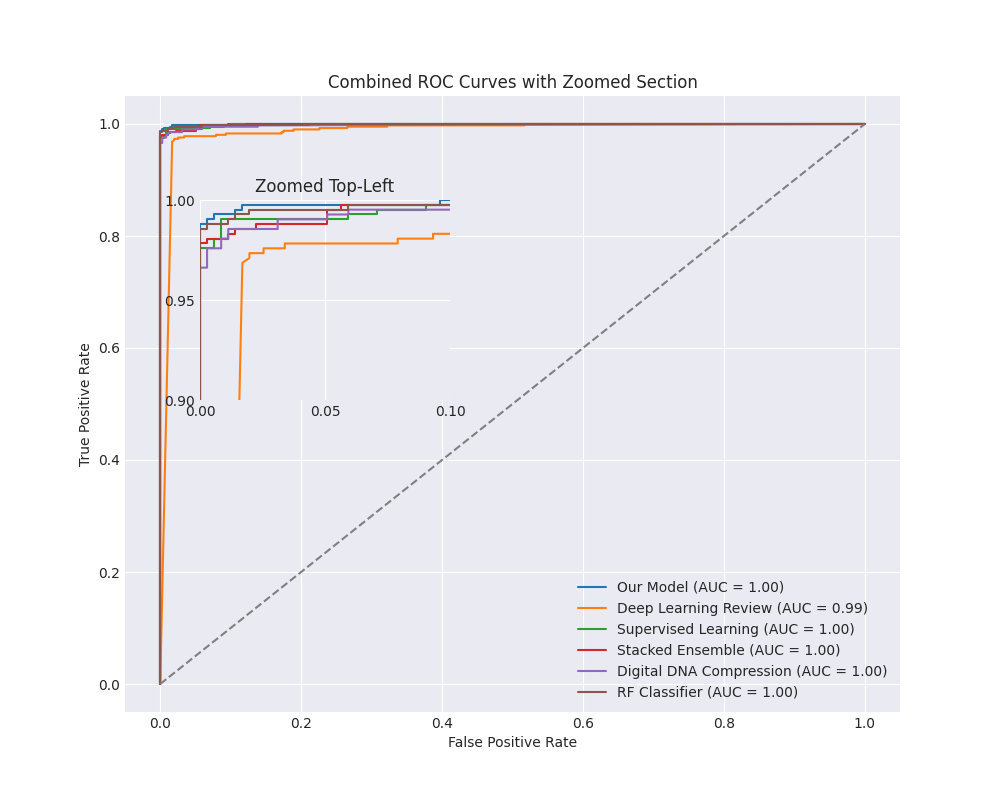}
    \caption{ROC curves comparing our model with other approaches}
    \label{fig:roc_curve}
\end{figure}

The ROC curve evaluates the model’s trade-off between True Positive Rate (TPR) and False Positive Rate (FPR). A higher AUC (Area Under the Curve) indicates stronger classification performance. Our model achieves an AUC close to 1.00 (Fig.~\ref{fig:roc_curve}), reflecting near-perfect discrimination between bot and human users. The closer the curve lies to the top-left corner, the better the model performs in minimizing false positives while maximizing true positives. This strong ROC profile reinforces the effectiveness of our approach and demonstrates its superiority over baseline methods.

\section{Conclusion}

In this work, we proposed a graph-based bot detection framework that effectively combines semantic tweet features with user-level metadata to construct a user–user interaction graph without relying on follower–following information. By leveraging BERT for textual feature extraction and GraphSAGE for relational reasoning, our model captures both individual behavior and cross-user patterns. Extensive experiments on multiple benchmark datasets, including Cresci-15, Cresci-17, and PAN, demonstrate the robustness and adaptability of our approach, particularly in scenarios where social relationship data is unavailable. The results confirm that integrating semantic, behavioral, and structural information leads to strong bot detection performance.

\bibliographystyle{IEEEtran}
\bibliography{reference}

\vspace{12pt}
\color{red}

\end{document}